\documentclass[aps,prl,twocolumn,groupedaddress]{revtex4}

\usepackage{natbib}
\usepackage{graphicx}
\usepackage{amsmath,amssymb}
\usepackage{ulem}

\begin{document}

\title{Scaling and anisotropy of magnetohydrodynamic turbulence in a strong mean magnetic field}
\author{Roland~Grappin}
\email[]{Roland.Grappin@obspm.fr}
\affiliation{LUTH, Observatoire de Paris
}

\author{Wolf-Christian~M\"{u}ller}
\email[]{Wolf.Mueller@ipp.mpg.de}
\affiliation{Max-Planck Institut f\"ur Plasmaphysik, 85748 Garching, Germany}
\date{\today}

\begin{abstract}
We present a new analysis of the anisotropic spectral energy distribution in
incompressible magnetohydrodynamic (MHD) turbulence permeated by a strong mean magnetic field.
The turbulent flow is generated by high-resolution pseudo-spectral direct numerical simulations with large-scale isotropic forcing.
Examining the radial energy distribution for various angles $\theta$ with respect to $\mathbf{B}_0$ reveals a specific structure which remains hidden when not taking axial symmetry with respect to $B_0$ into account.
For each direction, starting at the forced large-scales, the spectrum first exhibits an amplitude drop around a wavenumber $k_0$ which marks the start of a scaling range and goes on up to a dissipative wavenumber $k_d(\theta)$.
The 3D spectrum for $k \ge k_0$ is described by a single $\theta$-independent functional form
$F(k/k_d)$, 
the scaling law being the same in every direction.
The previous properties still hold when increasing the mean field from $B_0=5$ up to  $B_0=10 \ b_{rms}$, as well as when passing from resistive to ideal flows.
We conjecture that at fixed $B_0$ the direction-independent scaling regime is reached when increasing the Reynolds number above a threshold which raises with increasing $B_0$. Below that threshold critically balanced turbulence is expected.
\end{abstract}
\maketitle

It is known that in presence of a mean magnetic field assumed here to
point in the $z$-direction, $\mathbf{B}_0=B_0\hat{\mathbf{e}}_z$,
nonlinear interactions in incompressible magnetohydrodynamics 
(MHD) are weakened in the field-parallel direction.
The MHD approximation allows to describe the large-scale dynamics of astrophysical plasmas, i.e. ionized gases, like the interstellar medium or the solar corona.
Due to the  above-mentioned anisotropy, the nonlinear energy transfer in MHD turbulence proceeds preferably to larger perpendicular spatial wavenumbers 
\cite{Strauss:1976p6094, Montgomery:1981p6103,
Shebalin:1983p6056,Grappin:1986p707}.
Direct numerical simulations (DNS) show that the field-perpendicular
energy spectrum exhibits  self-similar inertial-range scaling in wavenumber 
$\sim k^{-m}$ with $m=5/3$
\citep{Cho:2000p6223,Cho:2002p5358} for weak to moderate $B_0$, or $m=3/2$
\citep{Maron:2001p5566,Muller:2005p705, Mininni:2007p24, Mason:2008p1} for strong $B_0$.

\citet{Iroshnikov:1963p9274} and \citet{Kraichnan:1965p9279} proposed the first theory of the effect of a mean magnetic field on incompressible MHD turbulence. 
They remarked that any flow can be decomposed into a sum of weakly interacting waves with different wavevectors $k$, the term weak interaction meaning that
the characteristic time of deformation of the waves is much longer than their periods.
This led to the prediction of a slow cascade, with a spectral slope $m=3/2$ different from the Kolmogorov prediction $m=5/3$.

This theory used an isotropic measure of the propagation time based on the modulus of the wave vector,
\begin{equation}
<\omega>^{-1} = <(\mathbf{k}.\mathbf{B}_0)^{-1}> \simeq k B_0^{-1}
\label{dispersion}
\end{equation}
ignoring deliberately the waves with wave vectors perpendicular (or almost perpendicular) to the mean field, for which the deformation time should clearly be smaller than their period, and hence the interaction strong.
This was criticized by \citet{Goldreich:1995p4882}, who denied the possibility of the previous weak cascade to occur, and argued that the perpendicular strong cascade leading to $k_\perp^{-5/3}$ should be the only one present. For a large enough mean field, the perpendicular cascade should thus be restricted to a thin subset around the $k_\perp$ axis in Fourier space, the subset becoming thinner when the mean field increases.

More precisely, the subdomain in ($k_\parallel$, $k_\perp$) space where the perpendicular cascade is believed to occur is defined by a critical balance \citep{Goldreich:1995p4882} between the characteristic time of nonlinear interaction, 
$\tau_{NL} \simeq (k_\perp u_\lambda)^{-1}$, and the Alfv\'en time 
$\tau_A \simeq (k_\parallel B_0)^{-1}$, 
the fluctuations becoming correlated along the guide field up to a distance 
$ \simeq B_0 \tau_{NL}$,
where $u_\lambda$ is the typical magnitude of fluctuations at the scale $\lambda \simeq 1/k_\perp$.
Assuming a scaling law in the perpendicular direction, and spectral transfer dominated by strong coupling, i.e., $\chi = \tau_A/\tau_{NL} \gtrsim 1$, one obtains
the 3D spectrum:
\begin{equation}
E_3(k_\parallel,k_\perp) = k_\perp^{-m-q-1} f(\chi)
\label{E3GS}
\end{equation}
where $\chi=k_0^{1-q} k_\perp^{q} k_\parallel^{-1} b_{rms}/B_0$, with 
$m=5/3$ and
$q = 2/3$,
$f(\chi) \simeq 1$ for $|\chi| \ge 1$ and $f(\chi)$ negligible for $|\chi| \ll 1$.

Eq.~(\ref{E3GS}) actually suffers from two limitations when the mean field $B_0$ is significantly larger than the magnetic fluctuation rms value. 
First, the spectral slope is observed to become $m=3/2$ \citep{Maron:2001p5566,Muller:2005p705, Mininni:2007p24, Mason:2008p1} instead of the strong cascade value $m=5/3$; 
second, the time-scales ratio $\chi=\tau_A/\tau_{NL}$ becomes significantly smaller than unity \citep{Bigot:2008p8863} in the excited part of the spectrum, showing that the strict critical balance condition $\chi=1$ is too restrictive to describe the anisotropy of the cascade, or, in other words, that the cascade is more extended in the oblique directions than predicted by the critical balance condition. 
This has led several authors to suggest modifications which either still assume that the anisotropy is dictated by the critical balance condition
 \citep{Boldyrev:2006p4917,Gogoberidze:2007p8855},  
or propose that the time-scales ratio $\chi$ decreases with increasing $B_0$ \citep{Galtier:2005p27}.
All these phenomenologies predict a spectral form different from Eq.~(\ref{E3GS}), with different spectral slopes in the perpendicular and parallel directions.

In the present paper, we analyze the angular spectrum and find by taking slices along the radial directions that a unique spectral form (and slope) holds in all directions, with the radial power-law range extent depending on the angle with the mean field. As a result, a significant portion of this spectrum lies in a domain where the time-scales ratio $\chi$ is sub-critical, that is, much smaller than unity.

This study focuses on representative states
of fully-developed turbulence permeated by a strong mean magnetic
field with $B_0=5 \ b_\mathrm{rms}$
from high-resolution direct numerical simulations of
quasi-stationary MHD turbulence forced at
large scales.  The forcing is realized by freezing all modes (velocity and magnetic field) with $k \le 2$ in an
energetically roughly isotropic and equipartitioned state.
The driving of magnetic energy could be realized physically by large-scale fluctuations of
electrical current although here it is mainly applied to achieve a state of approximate equipartion of kinetic
and magnetic energy. Decaying test simulations have confirmed that the forcing does not modify the results
presented in the following.
The dimensionless equations of resistive MHD formulated with the vorticity $\omega=\nabla\times\mathbf{v}$ and the magnetic field $\mathbf{b}$ are given by
\begin{eqnarray*}
\partial_t\ \omega &=&\nabla\times\left[\mathbf{v}\times\omega-\mathbf{b}\times
\left(\nabla\times\mathbf{b}\right) \right] + \mu\Delta\ \omega\,, \label{firstmhd}\\
\partial_t\mathbf{b}&=&\nabla\times\left(\mathbf{v}\times\mathbf{b}\right)+\eta\Delta\mathbf{b}\,,\\
\nabla\cdot\mathbf{v}&=&\nabla\cdot\mathbf{b}=0\,.
\end{eqnarray*}
The equations are solved by a standard pseudospectral method with spherical mode
truncation to alleviate aliasing errors.
The numerical resolution is $1024^2 \times 256$
collocations points with reduced resolution in the direction of $\mathbf{B}_0$
\citep{Muller:2005p705} and with kinematic viscosity $\mu$  and resistivity $\eta$ set
to $\mu=\eta=9\cdot 10^{-5}$.
The analyzed data is the temporal average of five snapshots of the three-dimensional Fourier
energy distribution taken equidistantly  within about four to five field-perpendicular large-scale
turnover times, $T_{0,\perp}$, of
quasi-stationary turbulence where
$T_{0,\gamma}=L_0/v_{\gamma\mathrm{rms}}=\pi/\langle v_\gamma^2\rangle^{3/2}
\int\mathrm{d}\mathbf{k'}\delta(k'_\gamma)|v_\gamma(\mathbf{k'})|^2$, $\gamma\in\{x,y,z\}$,
cf. \citep{Zikanov:2004p8800}, and
$v_\mathrm{rms}\approx b_\mathrm{rms}=1$ with
$T_{0,\perp}\approx 1.5$, $T_{0,\parallel}\approx 1.7$.
The normalized cross helicity $\rho=\langle \mathbf{v}\cdot\mathbf{b}\rangle
/(\langle v^2\rangle^{1/2}\langle b^2\rangle^{1/2})$ and the Alfv\'en ratio $\langle v^2\rangle/\langle b^2\rangle$
fluctuate around $23\%$ and $93\%$.

The three-dimensional (3D) energy spectrum, $E_3(k_x,k_y,k_z)$,
relates to the total energy
$E^\mathrm{tot}=\int d^3x(v^2+b^2)/2 =
\int d^3k E_3(k_x,k_y,k_z)$.

\begin{figure}
\includegraphics [width=\linewidth]{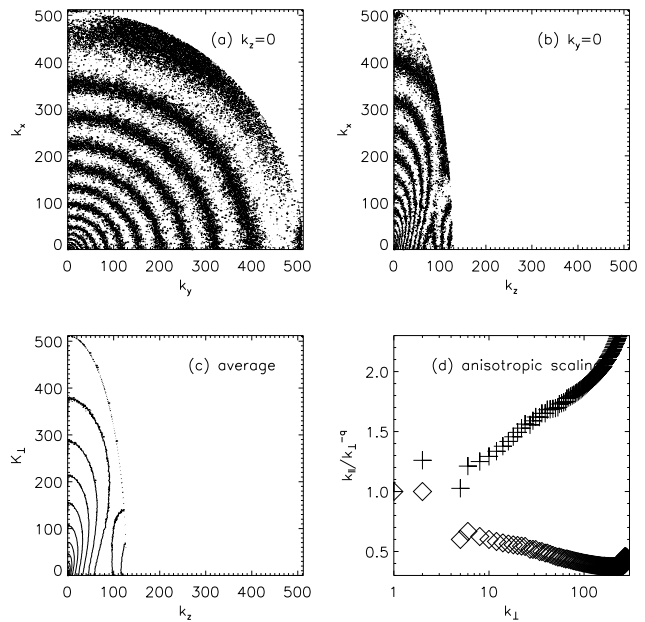}
\caption{Energy contour levels of three-dimensional spectral energy density $E_3(k_x,k_y,k_z)$ :
(a)plane $k_z=0$, (b) plane $k_y=0$,
(c) average of energy density over all planes containing $\mathbf{B}_0$.
(d) anisotropic scaling law between wave numbers $k_\perp$ and $k_\parallel$ (see text), with two compensated scalings: $k_\parallel(k_\perp)k_\perp^{-1}$ (diamonds) and $k_\parallel(k_\perp)k_\perp^{-2/3}$ (crosses).
}
\label{figA}
\end{figure}

Fig.~\ref{figA} shows contour levels of $E_3$ in two mutually orthogonal
planes containing the origin.
The field-perpendicular $k_x$-$k_y$ plane (Fig.~\ref{figA},a) displays an
isotropic energy distribution, as expected.
Anisotropy induced by $\mathbf{B}_0$ appears in 
planes containing the $B_0$ direction
 (Fig.~\ref{figA},b).

\begin{figure}[ht]
\begin{center}
\includegraphics [width=\linewidth]{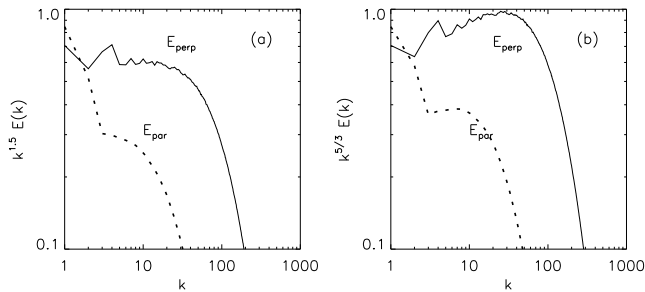}
\caption{
Plane-integrated one-dimensional perpendicular and parallel energy spectra compensated by (a) $k^{3/2}$ and (b) $k^{5/3}$, respectively. The symbol $k$ stands for the respective field-perpendicular and
field-parallel wavenumber
}
\label{figB}
\end{center}
\end{figure}

Spectral anisotropy is traditionally diagnosed by 1D spectra, e.g.,
$E_{\perp}(k_x) =\int d^3k' E_3 \delta(|k_x|-k_x') $
and
$E_{\parallel}(k_z) = \int d^3k' E_3 \delta(|k_z|-k_z')$.
These are shown compensated by $k^{3/2}$ (a) and by $k^{5/3}$ (b) in Fig.~\ref{figB}.
The perpendicular spectrum exhibits a power-law with $m=3/2$,
in agreement with previous works
\citep{Maron:2001p5566,Muller:2005p705, Mininni:2007p24, Mason:2008p1},
while the parallel spectrum does
not show any convincing scaling range.
An anisotropic scaling law is build from the previous 1D spectra by plotting
(Fig.~\ref{figA},d) the modes 
$(k_\perp,k_\parallel)$ sharing the same 1D energy density \citep{Bigot:2008p8863}. 
The anisotropy exponent $q$ is seen to lie between $q=1$ and $q=2/3$, which is also obtained in \citep{Bigot:2008p8863} for $B_0=5$.

While planar integration yields some information about anisotropy, it 
mixes all wavenumbers
perpendicular to the chosen direction and thus blurs the separation between
inertial and dissipative scales if the dissipative scale is not constant over the planar domain of integration. More importantly, no information on
intermediate directions between parallel and perpendicular is available.
Thus, spherical coordinates
$(k,\theta,\phi)$ with respect to the mean field axis along $\hat{\mathbf{e}}_z$ are considered.
As $E_3$ is isotropic in the azimuthal plane, the $\phi$-dependence of $E_3(k,\theta,\phi)$
is eliminated by averaging over $\phi\in [0,2\pi]$ which strongly decreases statistical noise
and yields
the $\phi$-averaged 3D spectrum
$E_3(k,\theta) = 1/(2\pi)\int d\phi E_3(k,\theta,\phi)$
whose isocontours in ($k_\parallel, k_\perp$) are shown in Fig.~\ref{figA},c.
We define the corresponding one-dimensional (1D) spectrum $E(k,\theta)$ as
$E(k,\theta) = k^2 E_3(k,\theta)$.
The total energy is thus
$E^\mathrm{tot} = 2\pi \int k^2 dk \int_{0}^{\pi/2} E_3(k,\theta) \sin(\theta) d\theta$.

\begin{figure}[ht]
\begin{center}
\includegraphics [width=\linewidth]{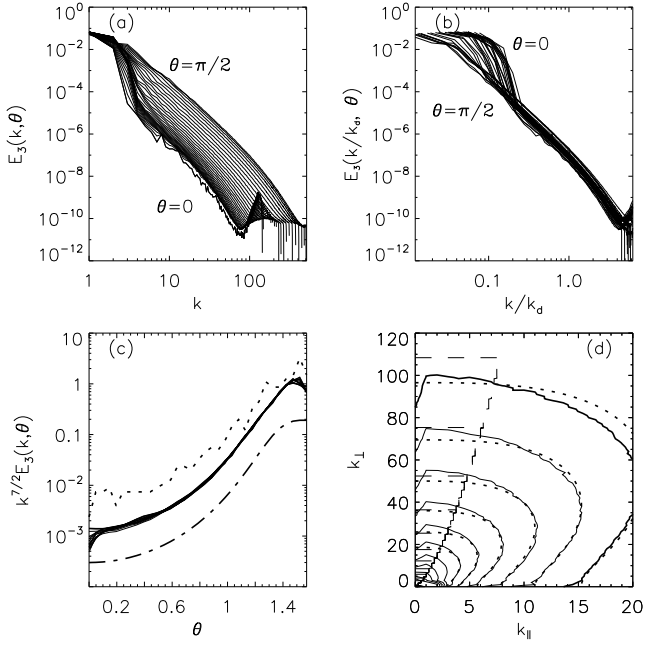}
\caption{Details of spectral properties:
(a) $E_3(k,\theta)$ for $\theta$ ranging
from $0$ to $\pi/2$,
(b) $E_3(k/k_d,\theta)$
(c) $k_d^{7/2}$ (dotted), $k^{7/2}E_3(k,\theta)$ for $8 \le k \le 20$ (solid),
Eq.(\ref{CV}) with $m=3/2$ (dotted-dashed)
(d) energy contour levels of $E_3(k_\parallel,k_\perp)$: simulation data (solid);
Eq.~\ref{E3GS} (dashed line, the oblique line tracing the boundary x=1 with $k_0=5$, $B_0=5$);
Eqs.~\ref{atheta1},\ref{CV} with $\beta=3$, $\varepsilon=0.158$ (dotted)
}
\label{figD}
\end{center}
\end{figure}

The properties of $E_3(k,\theta)$ are shown in
Fig.~\ref{figD}.  A scaling range is seen starting at $k_0 \simeq
4$--$8$ down to a dissipative wavenumber $k_d(\theta)$. The 1D scaling exponent
$m$
\begin{equation}
E_3(k,\theta)=A(\theta) k^{-m-2}
\label{atheta1}
\end{equation}
is \textit{independent} of $\theta$.
The anisotropy appears at fixed $k$ as a $\theta$-dependence of the spectral amplitude $A(\theta)$ and
of the dissipative wavenumber $k_d(\theta)$.
Normalizing the wavenumber by $k_d$ shows that the spectrum follows a single functional form
$F$
whatever $\theta$:
\begin{equation}
E_3(k,\theta)= F(k/k_d) = F_0 (k/k_d(\theta))^{-m-2}
\label{atheta2}
\end{equation}
where $F_0$ is a constant (the amplitude of the spectrum at the dissipative scale).
Note that the first equality holds also beyond the dissipative range, the second being valid for $k \le k_d$.
A corollary is that :
\begin{equation}
A(\theta) \propto k_d(\theta)^{m+2}\,.
\label{kd}
\end{equation}
Fig.~\ref{figD},a shows the self-similar wavenumber intervals of all $E_3(k,\theta)$ spectra,
starting in the range
$k_0 \simeq 4$--$8$ with a slight dependence on $\theta$.
The dissipative wavenumber $k_d(\theta)$ is estimated by locating the maximum of $k^2E(k)$ in each direction $\theta$:
varying $\theta$ from $\pi/2$ to $0$ leads to a drop of
$k_d(\theta)$ from about $100$ to $14$ while the spectral energy at fixed $k$
decreases in the inertial range from $1$ to $10^{-3}$.
Fig.~\ref{figD},b indicates that the spectral $\theta$-dependence can be nearly eliminated by normalizing with $k_d$ (Eq.~(\ref{atheta2})).
The relation between spectral amplitude $A(\theta)$ and $k_d(\theta)$ as given by Eq. (\ref{kd}) is confirmed by Fig.~\ref{figD},c which shows that
$k_d^{7/2}$ (dotted curve) closely follows $A(\theta)=E_3(k,\theta)k^{7/2}$ (Eq. (\ref{atheta1})),
with $k \le 20$ to eliminate the dissipative range in the parallel direction.

A simple model of anisotropic spectrum with a spectral exponent being the same in all directions has been proposed previously in the context of shell models of turbulence \cite{Carbone:1990p6061} as well as solar wind turbulence \cite{Tessein:2009p7225, Carbone:1995p9492}. 
It reads:
\begin{equation}
A(\theta) = (cos^2 \theta /\varepsilon^2+sin^2 \theta)^{-(1+m/2)}
\label{CV0}
\end{equation}

We tried to use this model to adjust the energy contours of our numerical simulations, and found that the global anisotropy between the perpendicular and parallel amplitude requires $\varepsilon = 0.158$. However, the model fails to reproduce correctly the detailed angular anisotropy, that is, the contours in oblique directions, because our energy contours differ much from ellipsoids, as is seen in Fig.~\ref{figD},d which shows a zoom of the energy contours as solid lines (the model of Eq.~(\ref{CV0}) would produce circular contours in this figure).
However we found that switching from the $2$ to $3$ for the exponents of the $sine$ and $cosine$ as:
\begin{equation}
A(\theta) = (cos^3 \theta /\varepsilon^2+sin^3 \theta)^{-(1+m/2)}
\label{CV}
\end{equation}
leads  to a reasonable good fit to the simulation results in all directions, as seen both in the dotted-dashed curve in Fig.~\ref{figD},c for the amplitude variation vs $\theta$ and in the dotted contours in the $(k_\parallel,k_\perp)$ plane in Fig.~\ref{figD},d.

The energy contours of the critical balance spectrum (Eq.~\ref{E3GS}, with $B_0=5$ and $k_0=5$) are also represented by dashed lines in Fig.~\ref{figD},d.
They isolate a small cone about the $k_\parallel$ axis in the whole plane, so excluding a 
large part of the angular structure of the true angular spectrum.

Note that the dissipative wavenumber is determined up to an error of about a factor two (due to errors in interpolating the spectrum), which leads to the noisy appearance of the curve of $k_d$ in Fig.~\ref{figD},c, to the finite thickness of the normalized spectra in Fig.~\ref{figD},b and to variations of about a factor 4 in the constant $F_0$ in Eq.~(\ref{atheta2}).

\begin{figure}[ht]
\begin{center}
\includegraphics [width=\linewidth]{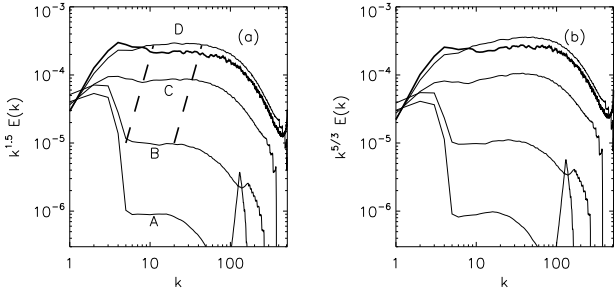}
\caption{Spectra averaged in four subsets of directions:
A: $14^0 \le \theta \le 39^0$; B: $42^0 \leq \theta \leq 65^0$;
C: $67^0 \le \theta \le 76^0$; D: $79^0 \leq \theta \leq 90^0$;
direction $\theta=90^0$ as thick curve.
(a): spectra compensated by $k^{3/2}$, the two oblique dashed lines
indicating roughly the inertial range; (b): spectra compensated by $k^{5/3}$
}
\label{figE}
\end{center}
\end{figure}

To determine the scaling exponent $m$ of $E(k,\theta)\sim k^{-m}$, the 1D spectra averaged over
four $\theta$-intervals  and compensated by
$k^{3/2}$ and $k^{5/3}$ are shown in Fig.~\ref{figE}.
The spectrum with $\theta=\pi/2$ is represented by the bold line.
The $m=3/2$-scaling is seen to be dominant, except possibly for group B which follows $m=5/3$ (Fig.~\ref{figE},a).
The extent of the $3/2$ inertial range is shown by oblique dashed lines;
it is bounded on the right by $k_d$, and on the left by an intermediate range which separates
the inertial from the forcing range.
The start of the inertial range is thus growing from $k\simeq 5$ to $10$ for $\theta\rightarrow\pi/2$.

Increasing the mean field up to $B_0=5 \sqrt{2}$ in test simulations (not shown) leads to further decrease of the power-law range in the parallel direction, with the perpendicular range increasing slightly, and the parallel range decreasing substantially, so that the ratio of both ranges varies with $B_0$ as
\begin{equation}
k_d(\pi/2)/k_d(0) =(A(\pi/2)/A(0))^{\frac{1}{m+2}} \simeq B_0
\label{ratio}
\end{equation}
Note that the fit by Eq.~(\ref{CV}) remains as good as in Fig.~\ref{figD},d after decreasing $\varepsilon$ by a factor $\sqrt 2$,
as expected since Eq.~(\ref{CV}) implies $k_d(\pi/2)/k_d(0) =1/\varepsilon$ hence from Eq.~(\ref{ratio}) $\varepsilon \simeq 1/B_0$.
Increasing again $B_0$ up to $10$ confirmed this trend, but the parallel range becomes too small in that case
to allow a good determination of the associated dissipative wavenumber. 
This precludes checking that the spectrum normalized by $k_d$ is angle-independent for small $\theta$.
This difficulty could be alleviated by increasing the numerical resolution which would allow increasing the Reynolds number. 
We choose instead below to compare with ideal MHD simulations with $512^3$ resolution and with $B_0=5$ and $10$.

\begin{figure}[ht]
\begin{center}
\includegraphics [width=\linewidth]{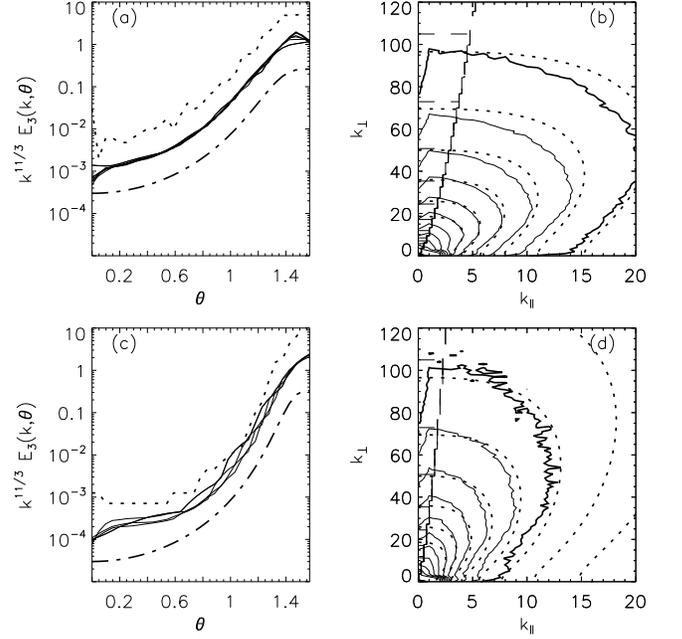}
\caption{Anisotropy in ideal runs; 
(a-b) $B_0=5$ (c-d) $B_0=10$.
(a)-(c): $k_d^{11/3}$ (dotted), $k^{11/3}E_3(k,\theta)$ (solid) (a) for $6 \le k \le 12$ 
and (b) for $5 \le k \le 8$,
Eq.(\ref{CV}) (dashed-dotted) with $m=5/3$, and (a) $\varepsilon=0.158$ (c) 
$\varepsilon=0.158/2$.
(b)-(d): energy contour levels of $E_3(k_\parallel,k_\perp)$: simulation data (solid);
Eq.~\ref{E3GS} (dashed line, the oblique line tracing the boundary $\chi=1$ with $k_0=5$, $B_0=5$ (b) and $B_0=10$ (d));
Eqs.~\ref{atheta1},\ref{CV} with $\beta=3$, $\varepsilon=0.158$ (b) and $\varepsilon=0.158/2$ (dotted)
}
\label{figideal}
\end{center}
\end{figure}
In the ideal case, Fourier space is divided in a large scale range presenting spectral properties close to those of a standard turbulent spectrum with dissipation, and a small scale range where the spectral slope increases, the latter scales playing the role of a dissipative range (cf. \citep{Cichowlas:2005p5664} in the hydrodynamic case).
The boundary between the two domains slowly shifts with time to ever larger scales. It can be identified with the dissipative wavenumber, thus determined here as the minimum of the 1D spectrum. 
The simulations are initialized with a quasi-stationary state of the resistive run and are continued without any dissipation and with the chosen $B_0$ in the same numerical setup until the energetically rising small scales begin to pollute the scaling region.
Choosing the appropriate time, power-law ranges in all directions can be properly identified even with a large field $B_0=10$. The resulting power-law is found to be now $m=5/3$ 
both with $B_0=5$ and $10$, contrary to the resistive runs.
This difference in scaling between the resistive and ideal runs might be attributed to a different role played by the bottlenkeck effect \cite{Beresnyak:2010p9399} in these two setups.
All reported findings, cf. Eqs.(\ref{atheta1}-\ref{ratio}), are however confirmed  when setting $m=5/3$, 
as seen in Fig.~\ref{figideal} which shows (cf. Fig.~\ref{figD},c) the anisotropy in two ideal runs with $B_0=5$ (a) and $B_0=10$ (b).

Let us come back to the resistive case. As already found in \citep{Bigot:2008p8863}, we find that the excited part of the $k_\parallel, k_\perp$ space is not restricted to regions where $\chi \ge 1$.
The important point is that the 3D-energy contours, and as well the boundary of the power-law range, ignore the iso-contours of $\chi$ as shown in Fig.~\ref{figD} (see also Fig~\ref{figideal}b,c): the form of the 3D-energy contours, as well as the angle-independent spectral slope, actually suggest an isotropic cascade, that is, a cascade along radial directions.
To test this idea, we define a $\theta$-dependent effective Reynolds $Re_0 \propto A(\theta)^{1/2}$ based on the energy density $A(\theta) k_0^{-m-2}$ at $k_0$.
For $\theta=0\rightarrow\pi/2$, $Re_0$ increases by a factor 30, 
while $k_d$ grows by a factor 10, as
\begin{equation}
k_d \propto Re_0^{\alpha}
\label{kdre}
\end{equation}
where $\alpha = 2/(m+2) \simeq 1/2$, $m$ being the 1D slope.
The exponent in Eq.~(\ref{kdre}) is substantially smaller than the value $\alpha = 3/4$ for $m=5/3$ (or $2/3$ in
the case $m=3/2$) obtained by equating the input flux at
$k_0$ and the dissipative flux $\epsilon \simeq \nu k_d^2 u(k_d)^2$).
This means that if $\theta$ increases from zero to $\pi/2$,
the inertial range increases more slowly than it would if the dissipative loss at small scales
would balance the input energy rate at the ($k_0 \simeq 8$) wavenumber which marks the
large-scale boundary of the inertial range.
Hence, for $\theta\rightarrow \pi/2$ the nonlinear radial energy flux must be depleted
while the contrary is true for the parallel directions.

By examining the solar wind turbulence, it has been shown \cite{Tessein:2009p7225} that the power-law index of the fluctuation spectra is independent of the angle between the wave vector and the mean interplanetary magnetic field, which is fully compatible with the results reported here from direct simulations.
Other studies of solar wind turbulence have however reached a different conclusion \citep{Horbury:2008p5853}.
The latter study used wavelet transforms, which allows to define parallel and perpendicular directions with respect to local averages of the magnetic field.
Indeed, it has been argued by \citep{Cho:2000p6223,Cho:2002p5358} that the critical balance phenomenon, and the associated spectral laws, (in particular the anisotropy index $q=2/3$ relating the perpendicular and parallel wave numbers) emerge only when considering, instead the mean field, the local average field.
In the work by \citep{Bigot:2008p8863} who consider as we do only the global mean field, the $q=2/3$ law appears clearly only when the mean field is large enough,
which can be explained by the fact that in this limit the local and mean field approach coincide.
According to this viewpoint, the results reported here should be a simple artifact of 
the fact that our frame is not attached to the local mean field, but to the global mean field.

However, while the effect could indeed appear for interplanetary turbulence where the fluctuation level is high, it is hardly the case here, since $B_0/b_{rms} = 5$ or $10$. 
Besides, it is remarkable that the fit by our anisotropy function in Fig.~\ref{figideal} is as good when $B_0=5 b_{rms}$ as when $B_0=10 b_{rms}$ which is an indication that the spectral properties of the turbulence are correctly revealed by using our method.

A possible way to reconcile both pictures is the following. 
It takes into account the fact that there is a second parameter, the Reynolds number.
Indeed, as the anisotropy increases with $B_0$, we find that the scaling range in the parallel direction decreases accordingly (Eq.~(\ref{ratio})), which implies that, at a fixed viscosity, the parallel power-law range disappears at even moderate $B_0$. 
In our case, $k_{d\perp} \simeq 100$ and $k_{d\parallel} \simeq 10$ when $B_0=5 b_{rms}$, while the start of the scaling range is $k_0 \simeq 5-8$, preventing to increase significantly $B_0$. 
The regime at high $B_0$ will thus depend on the Reynolds number ($Re$). 
Increasing $B_0$ at fixed perpendicular Re depletes the field-parallel cascade
and could ultimately lead to critically balanced turbulence.
If $B_0$ \textit{and} $Re$ are however large enough, e.g. under astrophysical conditions, allowing
for scaling in all directions then the properties described in this work are expected
to hold.
We thus propose direction-independent scaling for high $Re$ and critically balanced turbulence at low $Re$. 
The crossover $Re$-value is expected to increase with $B_0$.
A strong indication in this sense is found in \citep{Bigot:2008p8863} where the anisotropy scaling exponent $q$ is seen to be between 1 and $2/3$ at $B_0=5 b_{rms}$ as here (Fig. \ref{figA},d), while it begins to cluster around $2/3$ at $B_0/b_{rms} \ge 10$. Such a good agreement with $q=2/3$ is found already at $B_0/b_{rms} \simeq 1 $ in \citep{Cho:2000p6223} because of a moderate Reynolds number.

We have reported here two previously unknown properties of the MHD angular energy spectra: (i) its functional form is $A(\theta)f(k)$ (ii) the anisotropy function $A(\theta)$ is best expressed (Eq.~\ref{ratio}) as the ratio of the perpendicular over the parallel power-law range extent, which scales linearly with $B_0$ in the $B_0/b_{rms}=5,10$ interval considered here.
These results offer important tests for future theories of anisotropic turbulence. 

\begin{acknowledgments}
We thank G. Belmont, J. L\'eorat, and A. Busse for several fruitful discussions.
\end{acknowledgments}

\bibliography{grappin}

\end{document}